\documentclass[11pt,a4paper]{article}
\usepackage{amsmath}
\usepackage{graphicx}

\input{tcilatex}
\begin{document}

\title{THE INFLUENCE OF THE DISCRETIZED RASHBA SPIN-ORBIT \\
INTERACTION ON THE HARPER MODEL}
\author{ O. BORCHIN\, and E. PAPP\, \\
Department of Theoretical Physics, Faculty of Physics \\
West University of Timisoara, Bul. Vasile Parvan nr.4, RO-300223 \\
Timisoara, Romania \vspace{1mm}\\
E-mail: ovidiuborchin@yahoo.com \\
E-mail: erhardt\_papp\_2005@yahoo.com}
\date{}
\maketitle

\begin{abstract}
The movement of the electrons under the simultaneous influence of a scalar
periodic potential and of a uniform transversal magnetic field is described
by the well-known second order discrete Harper equation. This equation
originates from a two-dimensional energy dispersion law under the minimal
substitution. Here one deals with the Harper model under the additional
influence of the discretized spin orbit interaction. Converting the
spin-orbit interaction in terms of discrete derivatives opens the way for
analytical and numerical studies. One finds coupled equations for the spin
dependent wave functions, which leads to an appreciable alteration of the
nested energy subbands characterizing the self-similar structure of the
usual Harper spectrum. To this aim the transfer matrix method has been
applied to selected spin-up and spin-down wavefunctions. Accordingly, very
manifestations of spinfiltering and of spin correlations are accounted for.
Our energy-bands calculations show that the splitting effect implemented by
such wavefunctions is appreciable.
\end{abstract}

\begin{center}
recursive energy bands

(Received \today)
\end{center}

{\centering}

\section{INTRODUCTION}

\qquad Two-dimensional systems of electrons submitted to the simultaneous
influences of the periodic potential and of a homogeneous external magnetic
field have received great interest \cite{wilkins, azbely} in the last four
decades. Such studies led to the celebrated Harper equation \cite{wilkins}: 
\begin{equation}
{\large \psi }_{n+1}+{\large \psi }_{n-1}+2\Delta \cos (2\pi \frac{\phi }{%
\phi _{0}}n+k_{2}a){\large \psi }_{n}=E_{s}{\large \psi }_{n}  \label{1}
\end{equation}%
which has been extensively investigated. This equation originates from the
influence of the minimal subtitution on the energy dispersion law. Here $%
E_{s}=2E/E_{0}$ denotes the dimensionless energy, which comes from a single
band tight binding description with nearest neighbor hopping.

Further, $\Delta $ denotes the anisotropy parameter, while $\beta =\phi
/\phi _{0}$ stands for a commensurability parameter expressing the number of
flux quanta per unit cell. The magnetic flux and the flux quantum are
denoted by $\phi =Ba^{2}$ and $\phi _{0}=h/e$ respectively, while $k_{2}a$
stands for the Brillouin phase.

Equation (\ref{1}), is a second order discrete equation which has been
studied numerically in terms of the transfer matrix method by Hofstadter 
\cite{hofstad}. It has been found that the spectrum is characterized by a
nested band structure i.e. by the so called Hofstadter-butterfly. Besides
experimental relevance \cite{gerhardts}, the Harper-equation provides ideas
for useful applications concerning electrons on lattices under the influence
of a transversal magnetic fields $(0,0,B)$ \cite{papp}.

On the other hand, it has been demonstrated that the influence of the
spin-orbit interaction $(SOI)$ \cite{dressel,rash} produce sensible effects 
\cite{winkle}. Such results open the way of controlling\ the electron's spin
under the influence of an external electric field or of a gate voltage, too 
\cite{datta}.

In order to investigate the influence of the Rashba-$SOI$ ($RSOI)$ on the
Harper model, we shall supplement the Harper Hamiltonian with the
discretized version of the Rashba energy term: 
\begin{equation}
H_{so}=\frac{\Lambda }{\hbar }(\sigma _{x}p_{y}-\sigma _{y}p_{x})  \label{2}
\end{equation}%
where $\Lambda $ is the Rashba coupling parameter, $\sigma _{x}$ and $\sigma
_{y}$ are the Pauli matrices, while $p_{x}$ and $p_{y}$ stand for momentum
operators, as usual.

This later discretization has its own interest as it provides the
possibility to account for the influence of an underlying electric field in
a rather consistent manner. This differs, of course, from the
magneto-electric Harper equation discussed before \cite{munoz,nazare,kunold}%
. In this paper one deals with the numerical energy-bands realization of the
superposition of (1) and (2), now by accounting for selected wave functions.
Such solutions concern spin-\textsl{up} and spin-\textsl{down} wave
functions, respectively, but inter-related ones will also be considered.

The paper is organized as follows. The discretized $RSOI$ model as well as
the derivation of the coupled equations are presented in section $II$.
Numerical investigations of the coupled equations are done in section $III$
by resorting to some selected wave functions. First one accounts solely for
the influence of the spin-\textsl{up} and spin-\textsl{down} wave functions,
respectively, which is reminescent to spin-filtered systems. Next, one deals
with inter-related spin-\textsl{up} and spin-\textsl{down} wave functions.
The conclusions are presented in section $IV$.

{\centering}

\section{DERIVATION OF COUPLED EQUATIONS}

\qquad In order to implement the discretized version of $RSOI$ into the
Harper model, we have to resort to the modified interaction Hamiltonian%
\begin{equation}
\widetilde{H}_{so}=c\Lambda 
\begin{pmatrix}
0 & \partial _{1}-i\partial _{2}+eBx/\hbar \\ 
-\partial _{1}-i\partial _{2}+eBx/\hbar & 0%
\end{pmatrix}
\label{3}
\end{equation}%
where $e>0$, $\partial _{1}=\partial /\partial x$ and $\partial
_{2}=\partial /\partial y$. For this purpose one proceeds by applying the
minimal substitution to (\ref{2}), by using the Landau-gauge for which the
vector potential is given by $(0,Bx,0)$.

Next, one resorts to the factorization%
\begin{equation}
{\large \psi (x,y)=\exp (ik}_{2}y)%
\begin{pmatrix}
{\large \psi }_{\uparrow }{\large (x)} \\ 
{\large \psi }_{\downarrow }{\large (x)}%
\end{pmatrix}
\label{4}
\end{equation}%
where ${\large \psi }_{\uparrow }{\large (x)}$ and ${\large \psi }%
_{\downarrow }{\large (x)}$stands for spin-\textsl{up} and spin-\textsl{down}
wave functions, respectively. This leads to the $RSO$-Hamiltonian%
\begin{equation}
\widetilde{H}_{so}=c\Lambda 
\begin{pmatrix}
0 & D_{1}^{(+)}\text{ } \\ 
D_{1}^{(-)}\text{ } & 0%
\end{pmatrix}
\label{5}
\end{equation}%
where%
\begin{equation}
D_{1}^{(\pm )}=\pm \partial _{1}+k_{2}+\frac{e}{\hbar }Bx\text{ .}  \label{6}
\end{equation}

On the other hand, the Harper-equation is generated by the energy dispersion
law\qquad 
\begin{equation}
E_{disp}(k_{1},k_{2})=E_{0}(\cos k_{1}a+\Delta \cos k_{2}a)  \label{7}
\end{equation}%
under the influence of the Peierls substitution. This results in the
substitution rules $k_{1}\rightarrow -i\partial _{1}$ and $k_{2}\rightarrow
-i\partial _{2}+eBx/\hbar $. These rules are responsible for the onset\ of
the Harper-Hamiltonian such as given by%
\begin{equation}
\widetilde{H}_{Har}=\frac{E_{0}}{2}[2\cosh (a\partial _{1})+2\Delta \cos (%
\frac{aeBx}{\hbar }+k_{2})]\text{ .}  \label{8}
\end{equation}

One would than obtain the second-order discrete equation%
\begin{equation}
\frac{2}{E_{0}}\widetilde{H}_{Har}{\large \psi }_{\uparrow ,\downarrow }(x)=%
{\large \psi }_{\uparrow ,\downarrow }(x+a)+{\large \psi }_{\uparrow
,\downarrow }(x-a)+2\Delta \cos (\frac{ae}{\hbar }Bx+k_{2}a){\large \psi }%
_{\uparrow ,\downarrow }(x)  \label{9}
\end{equation}%
which reproduces equation (\ref{1}) in terms of the space-discretization $%
x=na$, where $n$ is an arbitrary integer. What then remains is to establish
the discretized version of equation (\ref{5}), by proceeding via%
\begin{equation}
\partial _{1}f(x)=\delta _{1}f(x)+\mathcal{O}(a^{2})  \label{10}
\end{equation}%
where $\delta _{1}$ stands for discrete derivative. It is understood that%
\begin{equation}
\delta _{1}f(x)=\frac{1}{a}\sinh (a\partial _{1})f(x)=\frac{1}{2a}({\large f}%
(x+a)-{\large f}(x-a))\text{ .}  \label{11}
\end{equation}

Accordingly, $\partial _{1}$ and $\delta _{1}$become identical to first $a$%
-order. This shows that the discretization one looks for, is provided by the
modified substitution rule%
\begin{equation}
D_{1}^{(\pm )}\rightarrow \pm \delta _{1}+k_{2}+2\pi n\frac{1}{a}\frac{\phi 
}{\phi _{0}}  \label{12}
\end{equation}%
which stands for the discrete counterpart of equation (\ref{6}). The
eigenvalue equation characterizing the total Hamiltonian is then given by%
\begin{equation}
H_{tot}{\large \psi }_{\uparrow ,\text{ }\downarrow }(x)=(\widetilde{H}%
_{Har}+\widetilde{H}_{so}){\large \psi }_{\uparrow ,\text{ }\downarrow }(x)=E%
{\large \psi }_{\uparrow ,\text{ }\downarrow }(x)  \label{13}
\end{equation}%
where $E$ denotes the total energy and where $\widetilde{H}_{so}$\ stands
for the discretized $RSO$-Hamiltonian. This results in the coupled equations%
\begin{eqnarray}
E_{s}{\large \varphi }_{\uparrow }(n) &=&{\large \varphi }_{\uparrow }(n+1)+%
{\large \varphi }_{\uparrow }(n-1)+2\Delta \cos (2\pi \beta n+k_{2}a){\large %
\varphi }_{\uparrow }(n)+  \notag \\
&&  \notag \\
&&+\frac{\Gamma }{2a}[{\large \varphi }_{\downarrow }(n+1)-{\large \varphi }%
_{\downarrow }(n-1)]+{\large \varphi }_{\downarrow }(n)\frac{\Gamma }{a}%
(k_{2}a+2\pi \beta n)  \label{14}
\end{eqnarray}%
with%
\begin{eqnarray}
E_{s}{\large \varphi }_{\downarrow }(n) &=&{\large \varphi }_{\downarrow
}(n+1)+{\large \varphi }_{\downarrow }(n-1)+2\Delta \cos (2\pi \beta
n+k_{2}a){\large \varphi }_{\downarrow }(n)+  \notag \\
&&  \notag \\
&&+\frac{\Gamma }{2a}[-{\large \varphi }_{\uparrow }(n+1)+{\large \varphi }%
_{\uparrow }(n-1)]+{\large \varphi }_{\uparrow }(n)\frac{\Gamma }{a}%
(k_{2}a+2\pi \beta n)  \label{15}
\end{eqnarray}%
where $\Gamma =-2\Lambda /E_{0}$ contains the Rashba coupling parameter. In
order to highlight the meaning of the coupled equations, we shall resort,
this time, to some quickly tractable simplifications.

{\centering}

\section{NUMERICAL INVESTIGATIONS}

\qquad At this point we have to realize that $E_{0}$ can be established in
terms of the effective mass $m^{\ast }$ of the electron as $E_{0}=-\hbar
^{2}/m^{\ast }a^{2}$. Using $m^{\ast }=0.067$m$_{e}$\ and $a=10$nm \cite%
{silber} then gives $E_{0}\simeq 11.37$meV. In addition one has $\Lambda
\simeq 5\cdot 10^{-11}$eVm \cite{catoixa, rowe}. A first simplification is
to assume that spin-down wave functions are zero ${\large \varphi }%
_{\downarrow }(n)=0$\textbf{,} which can be viewed as relying on spin filter
devices \cite{amado} or spin separation processes \cite{guo}.

This Ansatz\ leads to a $s=1$ spin-selection and $\theta _{2}=k_{2}a$, as%
\begin{equation}
{\large \varphi }_{\uparrow }(n+1)+{\large \varphi }_{\uparrow
}(n-1)=-2\Delta \cos (2\pi \beta n+\theta _{2})+E_{s}{\large \varphi }%
_{\uparrow }(n)  \label{16}
\end{equation}%
with%
\begin{equation}
{\large \varphi }_{\uparrow }(n+1)-{\large \varphi }_{\uparrow }(n-1)=2(2\pi
\beta n+\theta _{2}){\large \varphi }_{\uparrow }(n)  \label{17}
\end{equation}%
which proceed irrespective of $\Gamma $. Under such conditions we obtain%
\begin{equation}
{\large \varphi }_{\uparrow }(n+1)={\large \varphi }_{\uparrow }(n)[\frac{%
E_{s}}{{\small 2}}-\Delta \cos (2\pi \beta n+\theta _{2})+(2\pi \beta
n+\theta _{2})]  \label{18}
\end{equation}%
with%
\begin{equation}
{\large \varphi }_{\uparrow }(n-1)={\large \varphi }_{\uparrow }(n)[\frac{%
E_{s}}{{\small 2}}-\Delta \cos (2\pi \beta n+\theta _{2})-(2\pi \beta
n+\theta _{2})]  \label{19}
\end{equation}%
so that%
\begin{equation}
\frac{{\large \varphi }_{\uparrow }(n-1)}{{\large \varphi }_{\uparrow }(n)}=%
\frac{E_{s}-2\Delta \cos (2\pi \beta n+\theta _{2})-2(2\pi \beta n+\theta
_{2})}{{\small 2}}\text{ \ }  \label{20}
\end{equation}%
and%
\begin{equation}
\frac{{\large \varphi }_{\uparrow }(n)}{{\large \varphi }_{\uparrow }(n+1)}=%
\frac{2}{E_{s}-2\Delta \cos (2\pi \beta n+\theta _{2})+2(2\pi \beta n+\theta
_{2})}\text{ .}  \label{21}
\end{equation}

Applying the standard transfer matrix method, one obtains\ 
\begin{equation}
\tbinom{{\large \varphi }_{\uparrow }(1)}{{\large \varphi }_{\uparrow }(2)}%
=T_{1}\tbinom{{\large \varphi }_{\uparrow }(0)}{{\large \varphi }_{\uparrow
}(1)}\text{ \ and \ }\tbinom{{\large \varphi }_{\uparrow }(2)}{{\large %
\varphi }_{\uparrow }(3)}=T_{2}T_{1}\tbinom{{\large \varphi }_{\uparrow }(0)%
}{{\large \varphi }_{\uparrow }(1)}  \label{22}
\end{equation}%
which can be generalized via%
\begin{equation}
\tbinom{{\large \varphi }_{\uparrow }(n)}{{\large \varphi }_{\uparrow }(n+1)}%
=T_{n-1}\cdots T_{2}T_{1}\tbinom{{\large \varphi }_{\uparrow }(0)}{{\large %
\varphi }_{\uparrow }(1)}\text{ .}  \label{23}
\end{equation}

This amounts to consider%
\begin{equation}
\tbinom{{\large \varphi }_{\uparrow }(n)}{{\large \varphi }_{\uparrow }(n+1)}%
=T_{n}\tbinom{{\large \varphi }_{\uparrow }(n-1)}{{\large \varphi }%
_{\uparrow }(n)}  \label{24}
\end{equation}%
where%
\begin{equation}
{\large T}_{n}=\left( 
\begin{array}{cc}
\frac{1}{\frac{E_{s}}{{\small 2}}\text{ }-\Delta \cos (2\pi \beta n+\theta
_{2})+(2\pi \beta n+\theta _{2})} & 0 \\ 
0 & \frac{E_{s}-2\Delta \cos (2\pi \beta n+2\pi \beta +\theta _{2})-2(2\pi
\beta n+2\pi \beta +\theta _{2})}{{\small 2}}%
\end{array}%
\right)  \label{25}
\end{equation}%
supposing that%
\begin{equation}
\left\vert Tr({\large T}_{n})\right\vert =\left\vert 2\Delta \cos (2\pi
\beta n+\theta _{2})\right\vert \leq 2\Delta \text{ .}  \label{26}
\end{equation}

Note that $\Delta =1$ case, which will be assumed hereafter, represents the
critical point of a metal-insulator transition. We are by now in position to
perform numerical studies in accord with (\ref{24})-(\ref{25}). One realizes
that this time, the usual symmetries of the Harper-spectrum like such as the
exact energy reflection symmetry $E\rightarrow -E$ and the $\beta $%
-symmetry, i.e. the symmetry under $\beta =1/2\rightarrow \beta =1-1/u$ ($%
u=1,2,3,4,...$), cease to be fulfilled.

The understanding is that by virtue of the $s=1$-selection one deals with a
different experimental situation. The flux dependence of the energy bands
when $s=1$ (i.e. $\gamma =0$) is displayed in $Fig.1$, as illustrated by the
bands located in the upper half plane. However, we can also select $\varphi
_{\uparrow }=0$ (instead of $\varphi _{\downarrow }=0$), in which case one
deals with $s=-1$ (instead of $s=1$). 
\begin{equation*}
\FRAME{itbpF}{2.2957in}{2.2957in}{0in}{}{}{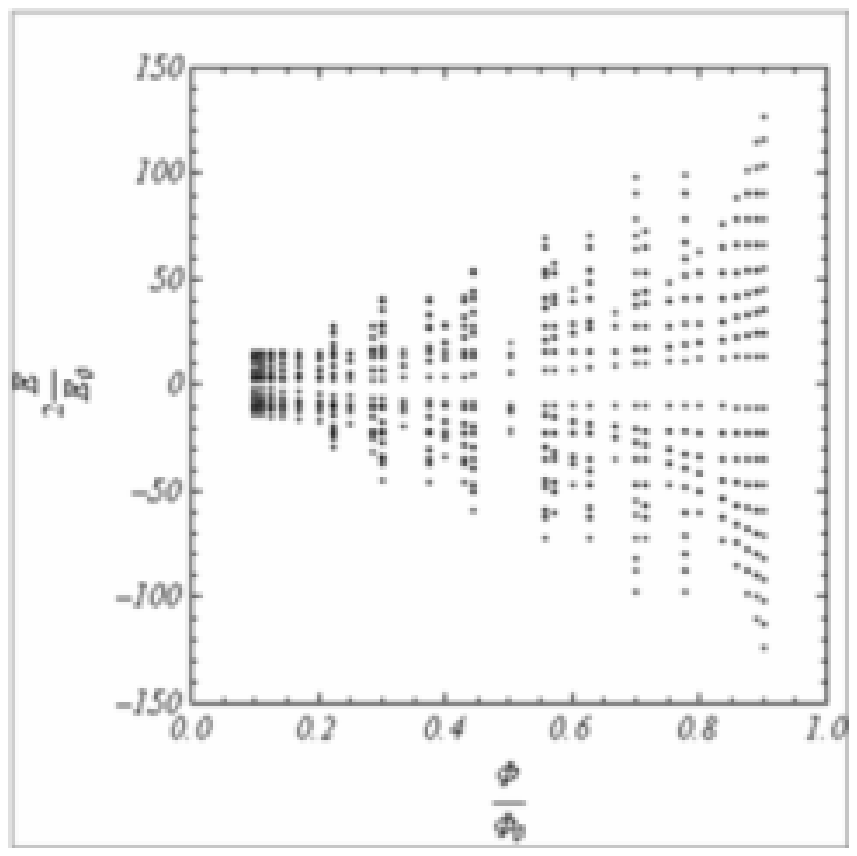}{\special{language
"Scientific Word";type "GRAPHIC";display "USEDEF";valid_file "F";width
2.2957in;height 2.2957in;depth 0in;original-width 3.3198in;original-height
3.3198in;cropleft "0.0078";croptop "0.9921";cropright "0.9921";cropbottom
"0.0078";filename 'fig.1.eps';file-properties "XNPEU";}}
\end{equation*}

\begin{center}
$Fig.1$ \ The flux dependence of energy bands for $s=1$ (upper

half plane) and $s=-1$ (lower half plane).
\end{center}

$^{{}}$

This time the energy-bands get located inside the lower half plane.

One remarks that bands become increasingly dispersed as the magnetix flux
increeases. This means that $\beta $- symmetry is completely lost. When the
flux approaches unity, the bands become equidistant, in accord with the
behavior of Landau levels. Such patterns can be gathered together as 
\begin{equation}
2\frac{E}{E_{0}}=sF_{\pm }(\beta ,k)  \label{27}
\end{equation}%
in which $s=1$ or $s=-1$, respectively. In addition there is $F_{\pm }(k)>0$
and%
\begin{equation}
F_{+}(\beta ,k)\approx F_{-}(\beta ,k)  \label{28}
\end{equation}%
which is responsible for energy reflection symmetry. The next approximation
is to consider inter-related spin-\textsl{up} and spin-\textsl{down} wave
functions say ${\large \varphi }_{\downarrow }(n)=$ $\gamma {\large \varphi }%
_{\uparrow }(n)$, where $\gamma $ is a real parameter. 
\begin{equation*}
\begin{array}{cc}
\FRAME{itbpF}{2.2957in}{2.2957in}{0in}{}{}{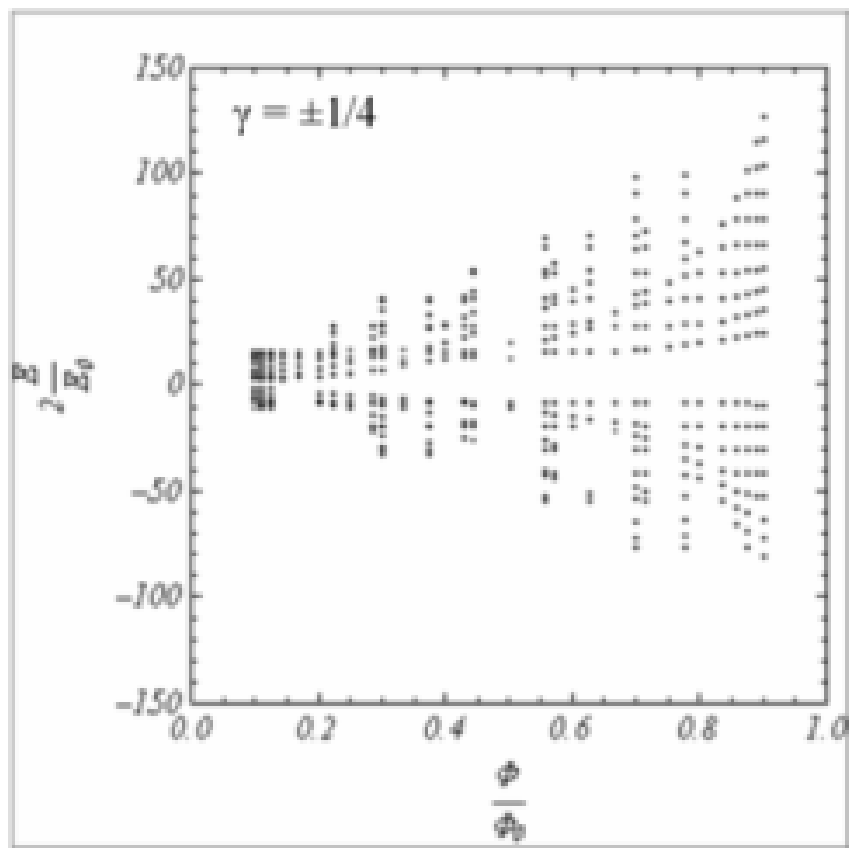}{\special{language
"Scientific Word";type "GRAPHIC";display "USEDEF";valid_file "F";width
2.2957in;height 2.2957in;depth 0in;original-width 3.3198in;original-height
3.3198in;cropleft "0.0078";croptop "0.9921";cropright "0.9921";cropbottom
"0.0078";filename 'fig.2.eps';file-properties "XNPEU";}} & \FRAME{itbpF}{%
2.2957in}{2.2957in}{0in}{}{}{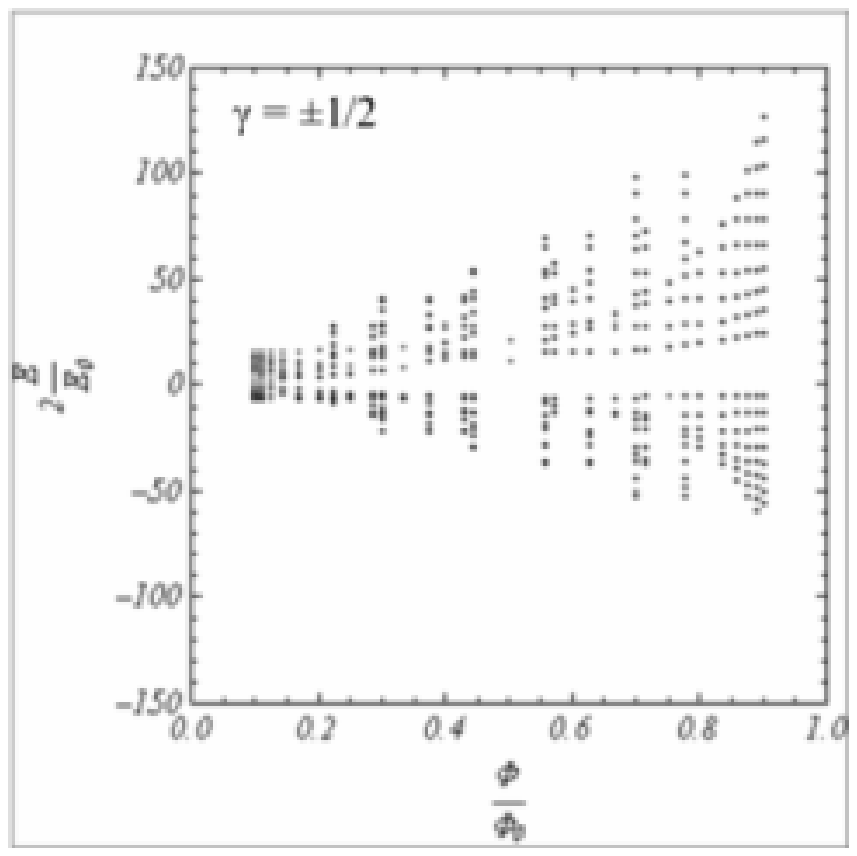}{\special{language "Scientific
Word";type "GRAPHIC";display "USEDEF";valid_file "F";width 2.2957in;height
2.2957in;depth 0in;original-width 3.3198in;original-height 3.3198in;cropleft
"0.0078";croptop "0.9921";cropright "0.9921";cropbottom "0.0078";filename
'fig.3.eps';file-properties "XNPEU";}} \\ 
\FRAME{itbpF}{2.2957in}{2.2957in}{0in}{}{}{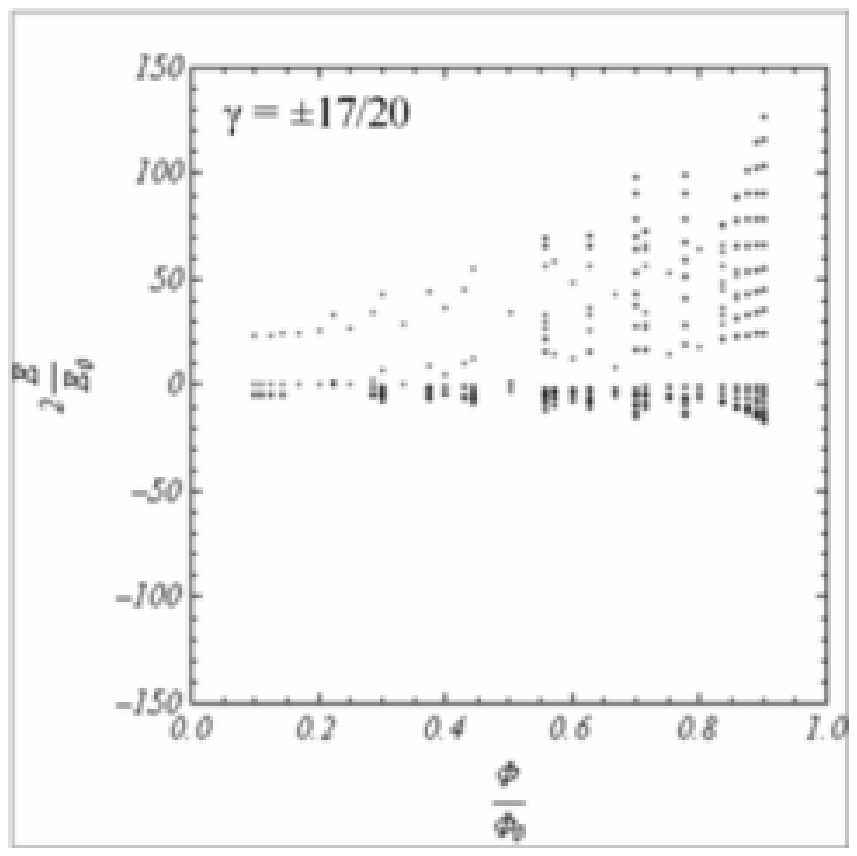}{\special{language
"Scientific Word";type "GRAPHIC";display "USEDEF";valid_file "F";width
2.2957in;height 2.2957in;depth 0in;original-width 3.3198in;original-height
3.3198in;cropleft "0.0078";croptop "0.9921";cropright "0.9921";cropbottom
"0.0078";filename 'fig.4.eps';file-properties "XNPEU";}} & \FRAME{itbpF}{%
2.2957in}{2.2957in}{0in}{}{}{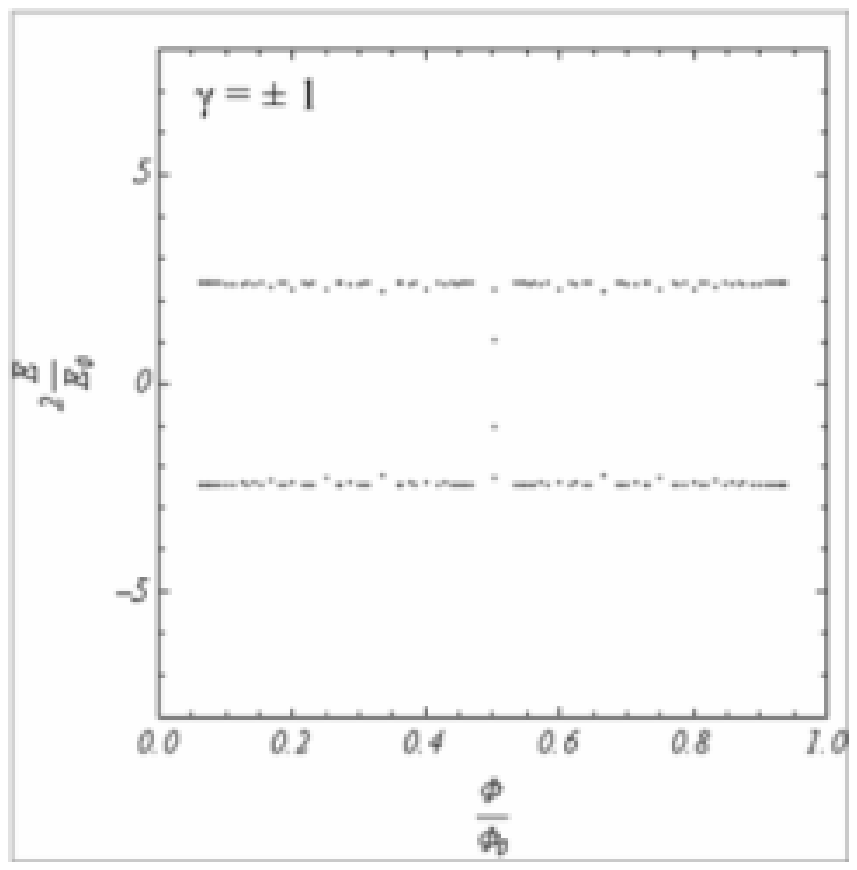}{\special{language "Scientific
Word";type "GRAPHIC";display "USEDEF";valid_file "F";width 2.2957in;height
2.2957in;depth 0in;original-width 3.3173in;original-height 3.3736in;cropleft
"0.0079";croptop "0.9921";cropright "0.9920";cropbottom "0.0078";filename
'fig.5.eps';file-properties "XNPEU";}}%
\end{array}%
\end{equation*}%
$\ \ \ \ \ $

\begin{center}
$Fig.2$ \ The flux dependence of energy bands for subunitary values of the

$\gamma $-parameter, like $\gamma =\pm 1/4,\pm 1/2,\pm 17/20$ and $\gamma
\pm 1$.
\end{center}

$^{{}}$

One would then obtain%
\begin{equation*}
\lbrack E_{s}-2\Delta \cos (2\pi \beta n+k_{2}a)-2\gamma \Gamma ^{\ast
}(k_{2}a+2\pi \beta n)]{\large \varphi }_{\uparrow }(n)=
\end{equation*}

\begin{equation}
=(1+\gamma \Gamma ^{\ast }){\large \varphi }_{\uparrow }(n+1)+(1-\gamma
\Gamma ^{\ast }){\large \varphi }_{\uparrow }(n-1)  \label{29}
\end{equation}%
with%
\begin{equation*}
\lbrack E_{s}\gamma -2\gamma \Delta \cos (2\pi \beta n+k_{2}a)-2\Gamma
^{\ast }(k_{2}a+2\pi \beta n)]{\large \varphi }_{\uparrow }(n)=
\end{equation*}

\begin{equation}
=(\gamma -\Gamma ^{\ast }){\large \varphi }_{\uparrow }(n+1)+(\gamma +\Gamma
^{\ast }){\large \varphi }_{\uparrow }(n-1)  \label{30}
\end{equation}%
by virtue of (\ref{14}) and (\ref{15}), respectively. 
\begin{eqnarray*}
&&\FRAME{itbpF}{2.2957in}{2.2957in}{0in}{}{}{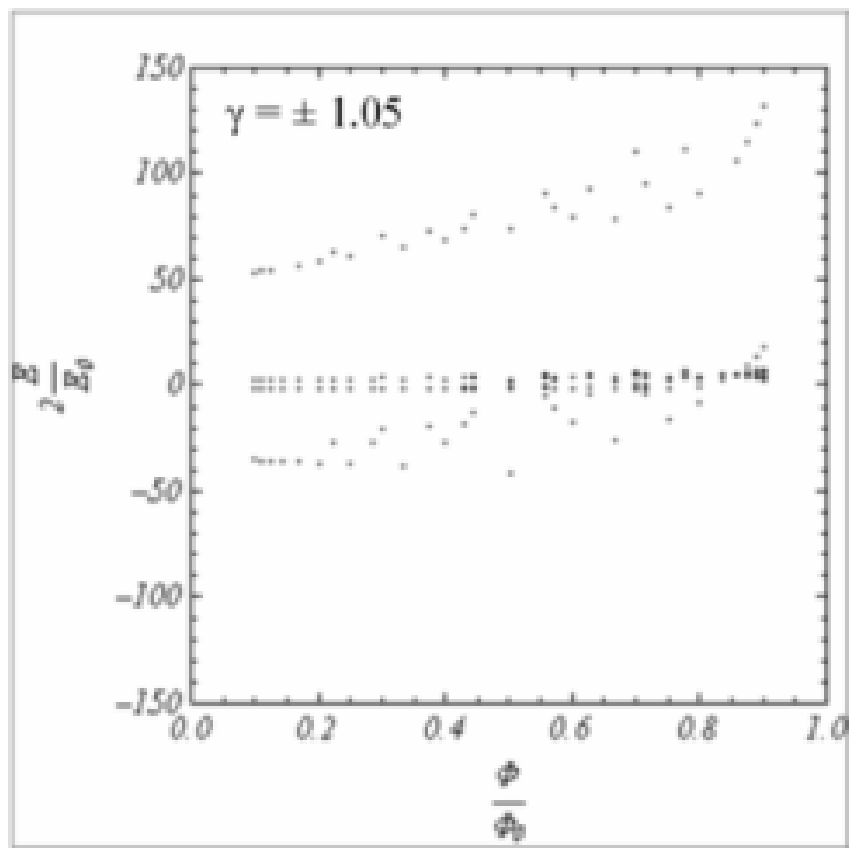}{\special{language
"Scientific Word";type "GRAPHIC";display "USEDEF";valid_file "F";width
2.2957in;height 2.2957in;depth 0in;original-width 3.3198in;original-height
3.3198in;cropleft "0.0078";croptop "0.9921";cropright "0.9921";cropbottom
"0.0078";filename 'fig.6.eps';file-properties "XNPEU";}}\FRAME{itbpF}{%
2.2957in}{2.2957in}{0in}{}{}{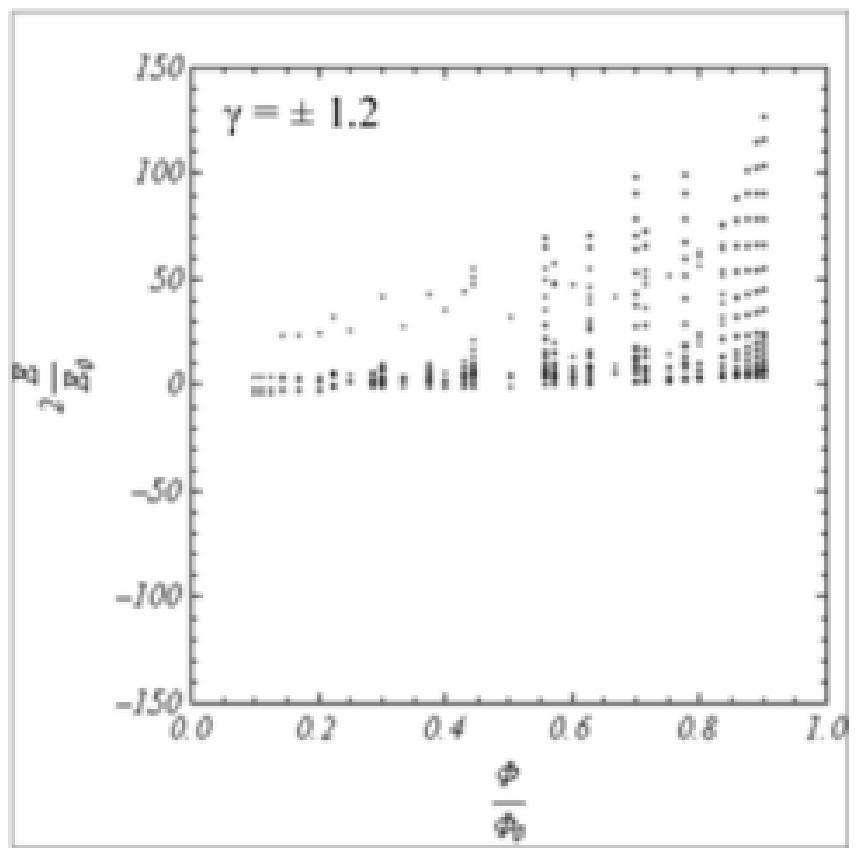}{\special{language "Scientific
Word";type "GRAPHIC";display "USEDEF";valid_file "F";width 2.2957in;height
2.2957in;depth 0in;original-width 3.3198in;original-height 3.3198in;cropleft
"0.0078";croptop "0.9921";cropright "0.9921";cropbottom "0.0078";filename
'fig.7.eps';file-properties "XNPEU";}} \\
&&\FRAME{itbpF}{2.2957in}{2.2957in}{0in}{}{}{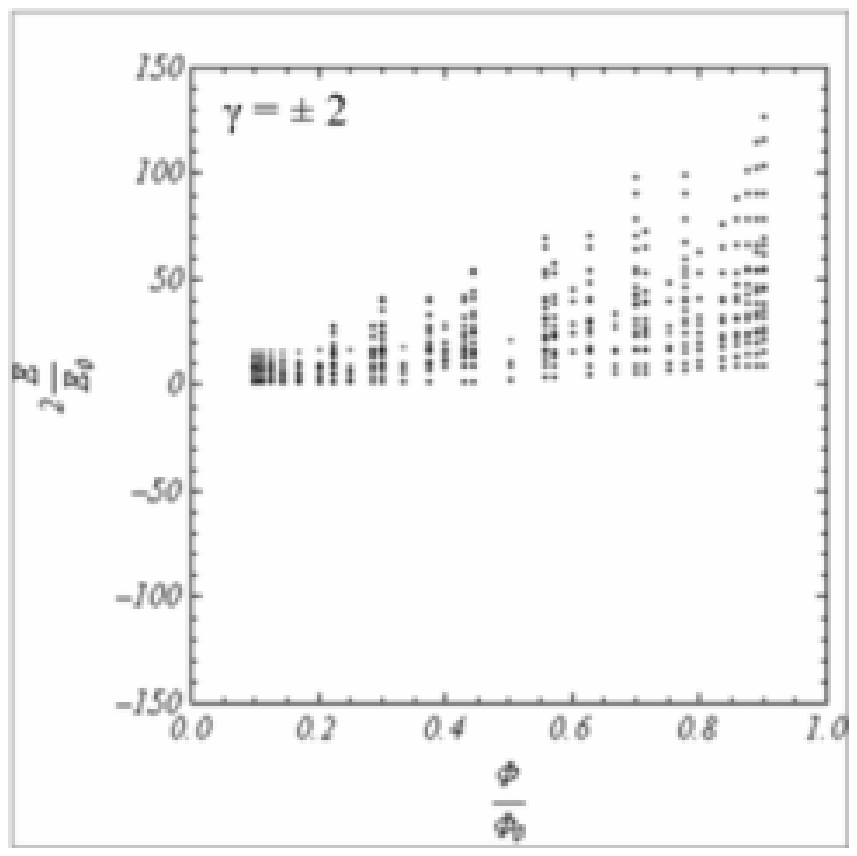}{\special{language
"Scientific Word";type "GRAPHIC";display "USEDEF";valid_file "F";width
2.2957in;height 2.2957in;depth 0in;original-width 3.3198in;original-height
3.3198in;cropleft "0.0078";croptop "0.9921";cropright "0.9921";cropbottom
"0.0078";filename 'fig.8.eps';file-properties "XNPEU";}}\FRAME{itbpF}{%
2.2957in}{2.2957in}{0in}{}{}{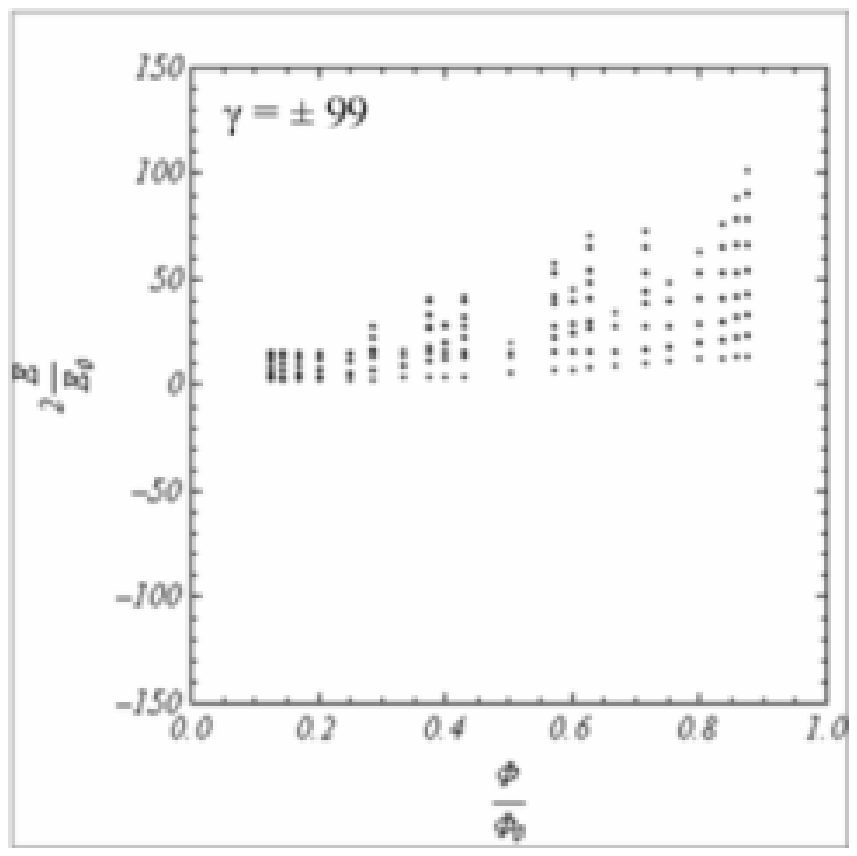}{\special{language "Scientific
Word";type "GRAPHIC";display "USEDEF";valid_file "F";width 2.2957in;height
2.2957in;depth 0in;original-width 3.3198in;original-height 3.3198in;cropleft
"0.0078";croptop "0.9921";cropright "0.9921";cropbottom "0.0078";filename
'fig.9.eps';file-properties "XNPEU";}}
\end{eqnarray*}

\begin{center}
$\qquad Fig.3$ \ The flux dependence of energy bands for

values of supraunitary modulus of the $\gamma $-parameter,

like $\gamma =\pm 1.05,\pm 1.2,\pm 2$ and $\gamma =\pm 99$.
\end{center}

$^{{}}$

These later equations lead in turn to%
\begin{equation}
\frac{{\large \varphi }_{\uparrow }(n-1)}{{\large \varphi }_{\uparrow }(n)}=%
\frac{(\gamma -\Gamma ^{\ast })E_{1}(n)-(1+\gamma \Gamma ^{\ast })E_{2}(n)}{%
(\gamma -\Gamma ^{\ast })(1-\gamma \Gamma ^{\ast })-(1+\gamma \Gamma ^{\ast
})(\gamma +\Gamma ^{\ast })}  \label{31}
\end{equation}%
and%
\begin{equation}
\frac{{\large \varphi }_{\uparrow }(n)}{{\large \varphi }_{\uparrow }(n+1)}=%
\frac{(\gamma +\Gamma ^{\ast })(1+\gamma \Gamma ^{\ast })-(1-\gamma \Gamma
^{\ast })(\gamma -\Gamma ^{\ast })}{(\gamma +\Gamma ^{\ast
})E_{1}(n)-(1-\gamma \Gamma ^{\ast })E_{2}(n)}  \label{32}
\end{equation}%
where $\Gamma ^{\ast }=\Gamma /2a$ so that $\Gamma ^{\ast }\simeq 7.043\cdot
10^{-20}$ is vanishingly small. In addition%
\begin{equation}
E_{1}(n)=E_{s}-2\Delta \cos (2\pi \beta n+k_{2}a)-2\gamma \Gamma ^{\ast
}(k_{2}a+2\pi \beta n)  \label{33}
\end{equation}%
and%
\begin{equation}
E_{2}(n)=E_{s}\gamma -2\gamma \Delta \cos (2\pi \beta n+k_{2}a)-2\Gamma
^{\ast }(k_{2}a+2\pi \beta n)  \label{34}
\end{equation}%
denotes the energies from (\ref{31}) and (\ref{32}). Using again the method
of the transfer matrix, one finds energy-band realization which are rahter
sensitive to $\gamma $. In addition we have to remark that by now $%
E_{s}=2E/E_{0}$ is an even function of $\gamma $. When $\left\vert \gamma
\right\vert <1$, the energy bands exhibit dispersions decreasing with $%
\left\vert \gamma \right\vert $, as shown in $Fig.2$ for $\gamma =\pm
1/4,\pm 1/2,\pm 17/20$ and $\gamma =\pm 1$. We have to remark that in this
latter case energy dispersion effect get inhibited, now by preserving both
energy reflection and $\beta $-symmetries. We have also to remark that the
plots displayed in $Fig.2$ for $\gamma =\pm 1$ reflect, within a reasonable
approximation, the onset of a two level configuration, say%
\begin{equation}
2\frac{E}{E_{0}}\simeq \pm 2.36  \label{35}
\end{equation}%
proceeding irrespective of the magnetic flux. Such configurations are of a
special interest for applications in the field of optoelectronics \cite%
{ragha, allen}. Next we found that supraunitary values of the $\gamma $
parameter lead to an asymmetric location in the flux dependence of energy
bands, as shown in $Fig.3$ for $\gamma =\pm 1.05,\pm 1.2,\pm 2$ and $\gamma
\pm 99$. One sees that in the later two cases the energies exhibit positive
values, only.

{\centering}

\section{CONCLUSION}

\qquad In this paper we have investigated both theoretically and numerically
the competition effects between the Harper model and the discretized $RSOI$.

Coupled equations for the spin dependent wave functions have been
established in some particular cases. Unlike the Harper spectrum, which, is
characterized by exact symmetries mentioned before, the presence of the $%
RSOI $ leads to dispersion effects in the flux dependence of the energy
bands, excepting the $\gamma =\pm 1$ plots in $Fig.2$.

Using the transfer matrix technique, we found that the $RSOI$ produces
dispersive effects in the energy dependence on the magnetic flux. Such
effects get visualized by a sensible alteration of the rather symmetrical
nested energy sub-bands characterizing the self-similar structure of the
usual Harper spectrum. We then have to realize that, excepting of course the
special $\gamma =\pm 1$ case mentioned above, both energy-reflection and $%
\beta $-symmetries are lost. Our first but restrictive approximation is to
account only for the influence of spin-\textsl{down} and spin-\textsl{up}
wave functions, respectively. The modified energy bands established in this
manner are displayed in $Fig.1$.

Our next approximation is to account for inter-related spin-\textsl{up} and
spin-\textsl{down} wave functions. Now we found that energy bands exhibit
again visible modifications, as shown in $Fig.2$ and $Fig.3$.

In other words we found that supplementing the Harper-Hamiltonian with the
discretized $RSOI$ leads to significant modifications of energy bands, which
serves to a better understanding of the electronic structure. This opens the
way to further applications concerning the influence of the spin-orbit
effect, with a special emphasis of two-level realizations mentioned above
which are of interest in the field of optoelectronics. More general wave
functions can also be considered, but such goals go beyond the immediate
scope of this paper.

$_{{}}$

{\large ACKNOWLEDGMENTS}

The authors are grateful to the referee for kind comments.

$_{{}}$

{\centering}


\begin{thebibliography}{99}
\bibitem{wilkins} M. Wilkinson, Proc. Roy. Soc. Lond., A403, 153 (1986).

\bibitem{azbely} M. Y. Azbel, Zh. Eksp. Teor. Fiz. 46, 939 (1964).

\bibitem{hofstad} D.R. Hofstadter, Phys. Rev. B14, 2239 (1976).

\bibitem{gerhardts} R.R. Gerhardts, D. Wein and U. Wulf, Phys. Rev. B 43,
5192 (1991).

\bibitem{papp} E. Papp and. C. Micu, \textsl{Low-Dimensional Nanoscale
Systems on Discrete Spaces}, World Scientific, Singapore, 2007.

\bibitem{dressel} G. Dresselhaus, Phys. Rev. 100, 580 (1955).

\bibitem{rash} E. I. Rashba, Fiz. Tverd. Tela (Leningrad) 2, (1960) 1224,
[Sov. Phys. Solid State 2, 1109 (1960)]; Y.A. Bychkov and E.I. Rashba, J.
Phys. C 17, 6039 (1984).

\bibitem{winkle} R. Winkler, \textsl{Spin-orbit coupling effects in
two-dimensional electron and hole systems}, Springer, Berlin, 2003.

\bibitem{datta} S. Datta and B. Das, Appl. Phys. Lett. 56, 665 (1990).

\bibitem{munoz} E. Mu\~{n}oz, Z. Barticevic and M. Pacheco, Phys. Rev. B 71,
165301 (2005).

\bibitem{nazare} H.N. Nazareno and P.E. de Brito. Phys. Rev. B 64, 045112
(2001).

\bibitem{kunold} A. Kunold and M. Torres, Phys. Rev. B 61, 9879 (2000).

\bibitem{silber} H. Silberbauer, J. Phys. Condens. Matter 4, 7355 (1992).

\bibitem{catoixa} X. Cartoix\`{a}, D.Z.-Y. Ting and T.C. McGill, Journal of
Computational Electronics 1, 141 (2002).

\bibitem{rowe} A.C.H. Rowe, J. Nehls, R.A. Stradling, R.S. Ferguson, Phys.
Rev. B 63, 201307(R) (2001).

\bibitem{amado} M. Amado, P. A. Orellana and F. Dom\'{\i}nguez-Adame,
Semicond. Sci. Technol. 21, 1764 (2006).

\bibitem{guo} Y. Guo and Xue-Zhen Dai, \textsl{Advanced Nanomaterials and
Nanodevices}, IUMRS-ICEM, Xi'an, China, 2002.

\bibitem{ragha} S. Raghavan, V. M. Kenkre, D. H. Dunlap, A. R. Bishop and M.
I. Salkola, Phys. Rev. A 54, R1781-1784 (1996).

\bibitem{allen} L. Allen and J. H. Eberly, \textsl{Optical Resonance and
Two-Level Atoms}, Wiley, New York, 1975).
\end{thebibliography}
\end{document}